\documentclass[twocolumn]{aastex631}

\begin{document}

\title{An Investigation Into The Selection and Colors of Little Red Dots and Active Galactic Nuclei}

\author[0000-0003-4565-8239] {Kevin N.\ Hainline}
\affiliation{Steward Observatory, University of Arizona, 933 N. Cherry Ave, Tucson, AZ 85721, USA}

\author[0000-0002-4985-3819] {Roberto Maiolino}
\affiliation{Kavli Institute for Cosmology, University of Cambridge, Madingley Road, Cambridge CB3 0HA, UK}
\affiliation{Cavendish Laboratory, University of Cambridge, 19 JJ Thomson Avenue, Cambridge CB3 0HE, UK}
\affiliation{Department of Physics and Astronomy, University College London, Gower Street, London WC1E 6BT, UK}

\author[0009-0003-7423-8660]{Ignas Juod\v{z}balis}
\affiliation{Kavli Institute for Cosmology, University of Cambridge, Madingley Road, Cambridge CB3 0HA, UK}
\affiliation{Cavendish Laboratory, University of Cambridge, 19 JJ Thomson Avenue, Cambridge CB3 0HE, UK}

\author[0000-0001-6010-6809]{Jan Scholtz}
\affiliation{Kavli Institute for Cosmology, University of Cambridge, Madingley Road, Cambridge CB3 0HA, UK}
\affiliation{Cavendish Laboratory, University of Cambridge, 19 JJ Thomson Avenue, Cambridge CB3 0HE, UK}

\author[0000-0003-4891-0794] {Hannah \"Ubler}
\affiliation{Kavli Institute for Cosmology, University of Cambridge, Madingley Road, Cambridge CB3 0HA, UK}
\affiliation{Cavendish Laboratory, University of Cambridge, 19 JJ Thomson Avenue, Cambridge CB3 0HE, UK}

\author[0000-0003-2388-8172] {Francesco D'Eugenio}
\affiliation{Kavli Institute for Cosmology, University of Cambridge, Madingley Road, Cambridge CB3 0HA, UK}
\affiliation{Cavendish Laboratory, University of Cambridge, 19 JJ Thomson Avenue, Cambridge CB3 0HE, UK}

\author[0000-0003-4337-6211] {Jakob M.\ Helton}
\affiliation{Steward Observatory, University of Arizona, 933 N. Cherry Ave, Tucson, AZ 85721, USA}

\author[0000-0001-6561-9443]{Yang Sun}
\affiliation{Steward Observatory, University of Arizona, 933 N. Cherry Ave, Tucson, AZ 85721, USA}

\author[0000-0002-4622-6617]{Fengwu Sun}
\affiliation{Center for Astrophysics $|$ Harvard \& Smithsonian, 60 Garden St., Cambridge MA 02138 USA}

\author[0000-0002-4271-0364] {Brant Robertson}
\affiliation{Department of Astronomy and Astrophysics, University of California, Santa Cruz, 1156 High Street, Santa Cruz CA 96054, USA}

\author[0000-0002-8224-4505] {Sandro Tacchella}
\affiliation{Kavli Institute for Cosmology, University of Cambridge, Madingley Road, Cambridge CB3 0HA, UK}
\affiliation{Cavendish Laboratory, University of Cambridge, 19 JJ Thomson Avenue, Cambridge CB3 0HE, UK}

\author[0000-0002-8651-9879] {Andrew J.\ Bunker}
\affiliation{Department of Physics, University of Oxford, Denys Wilkinson Building, Keble Road, Oxford OX1 3RH, UK}

\author[0000-0002-6719-380X] {Stefano Carniani}
\affiliation{Scuola Normale Superiore, Piazza dei Cavalieri 7, I-56126 Pisa, Italy}

\author[0000-0003-3458-2275] {Stephane Charlot}
\affiliation{Sorbonne Universit\'e, CNRS, UMR 7095, Institut d'Astrophysique de Paris, 98 bis bd Arago, 75014 Paris, France}

\author[0000-0002-9551-0534] {Emma Curtis-Lake}
\affiliation{Centre for Astrophysics Research, Department of Physics, Astronomy and Mathematics, University of Hertfordshire, Hatfield AL10 9AB, UK}

\author[0000-0003-1344-9475]{Eiichi Egami}
\affiliation{Steward Observatory, University of Arizona, 933 N. Cherry Ave, Tucson, AZ 85721, USA}

\author[0000-0002-9280-7594]{Benjamin D.\ Johnson}
\affiliation{Center for Astrophysics $|$ Harvard \& Smithsonian, 60 Garden St., Cambridge MA 02138 USA}

\author[0000-0001-6052-4234]{Xiaojing Lin}
\affiliation{Department of Astronomy, Tsinghua University, Beijing 100084, China}
\affiliation{Steward Observatory, University of Arizona, 933 N. Cherry Ave, Tucson, AZ 85721, USA}

\author[0000-0002-6221-1829]{Jianwei Lyu}
\affiliation{Steward Observatory, University of Arizona, 933 N. Cherry Ave, Tucson, AZ 85721, USA}

\author[0000-0003-4528-5639]{Pablo G. P\'erez-Gonz\'alez}
\affiliation{Centro de Astrobiolog\'ia (CAB), CSIC–INTA, Cra. de Ajalvir Km.~4, 28850- Torrej\'on de Ardoz, Madrid, Spain}

\author[0000-0002-5104-8245]{Pierluigi Rinaldi}
\affiliation{Steward Observatory, University of Arizona, 933 N. Cherry Ave, Tucson, AZ 85721, USA}

\author{Maddie S.\ Silcock}
\affiliation{Centre for Astrophysics Research, Department of Physics, Astronomy and Mathematics, University of Hertfordshire, Hatfield AL10 9AB, UK}

\author[0000-0001-8349-3055] {Giacomo Venturi}
\affiliation{Scuola Normale Superiore, Piazza dei Cavalieri 7, I-56126 Pisa, Italy}

\author[0000-0003-2919-7495] {Christina C.\ Williams}
\affiliation{NSF's National Optical-Infrared Astronomy Research Laboratory, 950 North Cherry Avenue, Tucson, AZ 85719, USA}

\author[0000-0001-9262-9997]{Christopher N.\ A.\ Willmer}
\affiliation{Steward Observatory, University of Arizona, 933 N. Cherry Ave, Tucson, AZ 85721, USA}

\author[0000-0002-4201-7367] {Chris Willott}
\affiliation{NRC Herzberg, 5071 West Saanich Rd, Victoria, BC V9E 2E7, Canada}

\author[0000-0002-1574-2045]{Junyu Zhang}
\affiliation{Steward Observatory, University of Arizona, 933 N. Cherry Ave, Tucson, AZ 85721, USA}

\author[0000-0003-3307-7525]{Yongda Zhu}
\affiliation{Steward Observatory, University of Arizona, 933 N. Cherry Ave, Tucson, AZ 85721, USA}
\begin{abstract}

Recently, a large number of compact sources at $z > 4$ with blue UV slopes and extremely red rest-frame optical slopes have been found in James Webb Space Telescope (JWST) extragalactic surveys. As a subsample of these sources, commonly called ``little red dots'' (LRDs), have been spectroscopically observed to host a broad-line active galactic nucleus (AGN), they have been the focus of multiple recent studies in an attempt to understand the origin of their UV and optical emission. Here, we assemble a sample of 123 LRDs from the literature along with spectroscopic and photometric JWST-identified samples of AGNs to compare their colors and spectral slopes. We find that while obscured AGNs at $z < 6$ have highly dissimilar colors to LRDs, unobscured AGNs at $z < 6$ span a wide range of colors, with only a subsample showing colors similar to LRDs. At $z > 6$, the majority of the unobscured AGNs that have been found in these samples are LRDs, but this may be related to the fact that these sources are at large bolometric luminosities. Because LRDs occupy a unique position in galaxy color space, they are more straightforward to target, and the large number of broad-line AGNs that do not have LRD colors and slopes are therefore underrepresented in many spectroscopic surveys because they are more difficult to pre-select. Current LRD selection techniques return a large and disparate population, including many sources having $2-5\mu$m colors impacted by emission line flux boosting in individual filters. 
\end{abstract}

\accepted{to the Astrophysical Journal, November 28, 2024}
\section{Introduction} \label{sec:intro}

The identification of large samples of active galactic nuclei (AGNs), especially at early cosmic times, has long been an observational challenge. These sources, which are powered by accretion onto the central supermassive black holes in their host galaxies, produce flux at wavelengths that span the full electromagnetic spectrum. However, methods for their selection are often biased towards finding sources that are significantly brighter compared to their host galaxy, or those that are less obscured by dust and gas \citep[see][for reviews]{padovani2017, hickox2018}. Because AGNs, and in particular those with broad hydrogen emission lines (also called Type I, or unobscured AGNs), can provide insight into the growth of supermassive black holes and the co-evolution of these black holes and the galaxies that host them, finding ways to target and collect large samples is crucial.

The sensitivity, resolution, and infrared wavelength coverage offered by the instruments on board the James Webb Space Telescope (JWST) have opened up our view of AGNs in the high-redshift Universe, extending our understanding of growing supermassive black holes to $z \sim 11$. Notably, these sources have luminosities significantly lower ($10^{44}$ erg s$^{-1}$ $< L_{\mathrm{bol}} < 10^{47}$  erg s$^{-1}$) than quasars at $z < 7$ identified before the launch of JWST, and are therefore at lower black hole masses (M$_{\mathrm{BH}} < 10^8 $M$_{\odot}$) or accretion rates ($< 0.5$ the Eddington rate) \citep{ubler2023, ubler2024, barro2024, harikane2023, kocevski2023, kocevski2024, larson2023, maiolino2023, maiolino2024, matthee2024, Kokorev23z8, juodzbalis2024a, ono2023, oesch2023, scholtz2023, Taylor2024}. Many of these sources were identified as either broad-line or narrow-line (Type II, or obscured) AGNs from features in their UV or optical spectra, and have been found in large spectroscopic surveys made with JWST/NIRSpec or the JWST/NIRCam grism. At $z > 5$, common methods for pre-selecting AGN candidates for follow-up, such as X-ray luminosity, mid-IR color, or radio emission, are difficult \citep[see e.g.][for discussions of these methods]{lyu2022,Maiolino24_Xrays,mazzolari2024,juodzbalis2024b}. Therefore, current populations of high-redshift AGNs found with JWST are likely incomplete.  

Among the AGN candidates discovered by JWST at high redshift, a surprising development has been the discovery of a large number of ultra-compact ($< 500$ pc) sources with blue rest-frame UV slopes ($\beta_{UV} < -0.37$) and extremely red rest-frame optical slopes ($\beta_{opt} > 0$) seen across multiple surveys. One of the earliest samples of these sources was assembled by \citet{matthee2024}, who identified AGNs with broad H$\alpha$ emission lines in galaxies selected from the Emission-line galaxies and Intergalactic Gas in the Epoch of Reionization
\citep[EIGER, PID 1243,][]{kashino2023} and First Reionization Epoch Spectroscopic COmplete Survey \citep[FRESCO, PID 1895,][]{oesch2023} NIRCam grism datasets. These authors found that the majority of these broad-line sources were red and compact, and they dubbed these objects ``little red dots'' (LRDs). Subsequent observations of photometrically-selected LRDs confirmed the broad line detections for a large fraction of these sources using NIRSpec \citep{greene2024, kocevski2024,Furtak24overmassive}. These LRDs were found at $z \sim 4 - 8$, and offered an intriguing look at the growth of supermassive black holes in the first few billion years of cosmic history.

Spurred by this discovery, multiple authors subsequently assembled samples of LRDs by searching through JWST/NIRCam extragalactic datasets first for compact objects and then by targeting their rest-frame colors through NIRCam color cuts \citep{akins2023, greene2024, barro2024, labbe2023b, kokorev2024, perezgonzalez2024} or through more complex cuts on spectral slope based on fits to the photometry \citep{kocevski2024}, targeting objects with a ``V-shaped'' UV-to-optical spectral energy distribution (SED). These disparate selection methods resulted in a diverse menagerie of recovered sources which included objects with red optical continua, but also sources only selected because of flux from strong optical emission lines. Brown dwarfs with effective temperatures $T_{\mathrm{eff}} < 1300$K have also been confused for LRDs as molecular absorption in their atmospheres can mimic the blue UV and red optical slopes \citep{burgasser2024, langeroodi2023, hainline2024}.

Understanding the origin of the UV and optical emission for LRDs has presented a mystery. In \citet{labbe2023a}, the authors fit a sample of LRDs with galaxy models and concluded that the optical slope may be evidence of these sources hosting massive ($> 10^{10} M_{\odot}$) stellar populations, well in excess of what would be expected at such early times. The presence of an AGN would help mitigate this issue, but LRD UV and optical SEDs are quite different from what is typically seen in AGNs at low redshift \citep{elvis1994, richards2006, hickox2018}. Multiple theories have arisen to explain what is being observed: the rest-frame UV emission could be a combination of galaxy stellar emission and scattered light from the central AGN, while the optical emission may be stellar continuum (with a Balmer break producing an observed spectral jump) or reddened AGN accretion disk continuum in combination with hot or warm dust \citep{kocevski2023, barro2024, labbe2023b, akins2023, perezgonzalez2024}. Other authors, such as \citet{li2024}, hypothesize that LRDs may be dust obscured but in a dusty medium lacking small-size grains, which produces a ``grey'' extinction in the UV with more significant reddening in the optical. More recently, \cite{Inayoshi_Maiolino_24} have suggested that, in most cases, the Balmer break observed in LRDs does not have a stellar origin and actually results from absorption from dense gas in the circumnuclear medium of AGNs, likely associated with clouds in the AGN broad line region or its surroundings.

Confusing the AGN interpretation is that observations of some LRDs in the mid-IR have shown that the rising optical slope in these sources flattens towards the rest-frame NIR ($1 - 3\mu$m), in conflict with AGN models \citep{williams2024, perezgonzalez2024, akins2024, wang2024}, although there are also cases showing near- to mid-IR excess typical of AGNs \citep{lyu2024,juodzbalis2024b}. In addition, the number counts of these sources are well in excess of the extrapolation of the quasar luminosity functions as derived from previous ground-based observations \citep{matsuoka2018, niida2020, he2024}. Finally, the lack of both X-ray emission and UV-optical variability for LRDs is also difficult to understand in light of the potential presence of an AGN in these sources  \citep{kokubo2024}. A study from \citet{baggen2024} suggested that the broad emission lines seen in LRDs may instead arise from the underlying kinematics of the hydrogen gas in extremely compact, massive environments.

Given the focus in the literature on LRDs and the origins of their emission, it is important to put their selection and colors in context within the samples of AGNs selected in other ways. While \citet{greene2024} found broad-line AGN fractions in their LRD sample of $>50\%$, one might wonder about the inverse: the overall broad-line LRD fraction of the whole population of AGNs. Do LRDs represent a dominant mode of AGN accretion at high-redshift, or do these sources differentiate themselves by virtue of ease of discovery? Given the variety of LRD selection methods, and even definitions of an LRD, that have appeared in the literature, what conclusions can be drawn about such a disparate population of sources? 

To that end, we have collected a large sample of LRDs and AGNs with photometry measured using uniform data reduction across the well-studied JWST Advanced Deep Extragalactic Survey (JADES)\footnote{https://jades-survey.github.io/} \citep{eisenstein2023} and The Cosmic Evolution Early Release Science (CEERS)\footnote{https://ceers.github.io/} \citep{finkelstein2022} deep fields. We combine our sample of LRDs selected photometrically and from fits to the photometry with broad- and narrow-line AGNs, as well as SED-derived AGNs across both fields. We explore the colors, compactness, UV and optical slopes, and bolometric luminosities for these sources compared to LRDs and find that the bulk of lower-luminosity AGN activity has not been probed yet by current surveys. 

In this paper we describe our measured GOODS-S, GOODS-N and EGS photometry, and the full set of LRDs, Type I and Type II AGNs we assembled in Section \ref{sec:observations}. We discuss the common methods for selecting LRDs from the literature, and show where the assembled sources in our sample live in color and slope space in Section \ref{sec:selection}. We discuss the implications for the colors and spectral slopes of the assembled samples in Section \ref{sec:results}. Finally, we discuss both the selection of LRDs and broad emission lines in LRDs in Section \ref{sec:discussion}, and conclude in Section \ref{sec:conclusions}. We assume the \citet{planck2020} cosmology, with $H_0 = 67.4$ km s$^{-1}$ Mpc$^{-1}$, $\Omega_{\mathrm{M}} = 0.315$ and $\Omega_\Lambda = 0.685$. All magnitudes are provided using the AB magnitude system \citep{oke1974, oke1983}. 

\section{Observations and Galaxy Samples} \label{sec:observations}

To explore the rest-frame UV and optical colors of LRDs and active galaxies, we looked at sources selected from the Great Observatories Origins Deep Survey Southern (GOODS-S, R.A. = 53.126 deg, Dec = -27.802 deg) and Northern (GOODS-N, R.A. = 189.229, Dec = +62.238 deg) regions \citep{giavalisco2004}, along with the Extended Groth Strip \citep[EGS,][R.A. = 214.25, Dec = +52.5 deg]{rhodes2000, davis2007}. Both of these regions have been imaged in part with JWST/NIRCam as part of the JADES and CEERS surveys. The depths and filters used in these surveys are ideal for disentangling the colors of both obscured and unobscured active galaxies and LRDs. We employ photometry derived by the JADES team for both the JADES GOODS-S, JADES GOODS-N and CEERS EGS NIRCam data to aid in the comparison in this study. 

The JADES data used in this paper includes all GOODS-S and GOODS-N observations taken as of January 2024, and the observations and reduction of these data is described in \citet{eisenstein2023}. We use JADES NIRCam observations taken as part of JWST PID 1180, 1181 (PI: Eisenstein) PID 1210, 1286 (PI: Ferruit), and PID 1287 (PI: Isaak). We also include NIRCam observations in GOODS-S and GOODS-N from the First Reionization Epoch Spectroscopic COmplete Survey \citep[FRESCO, PID 1895;][]{oesch2023}, as well as the JWST Extragalactic Medium-band Survey \citep[JEMS][]{williams2023} and JADES Origins Field \citep{eisenstein2023b}. We also include the deep NIRCam GOODS-S data from the Next Generation Deep Extragalactic Exploratory Public (NGDEEP, PID: 2079, PIs: S. Finkelstein, Papovich and Pirzkal) survey. For CEERS EGS, we use NIRCam data from ERS PID 1345 (PI: Finkelstein). For this work, we primarily focus on the NIRCam bands that are in common between JADES and CEERS: F115W, F150W, F200W, F277W, F356W, F410M, and F444W filters. Later, we will discuss how other NIRCam medium band filters that have been used to observe subsamples of sources in JADES are helpful: F162M, F182M, F210M, F250M, F300M, F430M, F460M, and F480M.

We include Hubble Space Telescope ACS mosaics from GOODS-S, GOODS-N, and the EGS to supplement the observations at shorter wavelengths. We use the HST/ACS mosaics from the Hubble Legacy Fields (HLF) v2.0 for GOODS-S and v2.5 for GOODS-N \citep{illingworth2013, whitaker2019}. We additionally use HST/ACS mosaics that cover the EGS (A. Koekemoer, private comm). For the present analysis, we only focus on data taken with the HST/ACS filters F606W and F814W, as these are the ACS filters available across both JADES and CEERS that are used in selecting LRDs.

The data for all four (JADES GOODS-S, JADES GOODS-N, NGDEEP GOODS-S and CEERS EGS) fields were reduced following the methodology described in \citet{robertson2023}, \citet{tacchella2023}, and \citet{eisenstein2023}. We primarily employ fluxes derived using 0.2$^{\prime\prime}$ diameter circular apertures, although for exploring compactness, we also use 0.5$^{\prime\prime}$ diameter circular apertures. At $z = 2$, 0.2$^{\prime\prime}$ - 0.5$^{\prime\prime}$ corresponds to 1.7 - 4.3 kpc, and at $z = 12$, 0.2$^{\prime\prime}$ - 0.5$^{\prime\prime}$ corresponds to 0.7 - 1.9 kpc. The 0.2$^{\prime\prime}$ apertures were chosen to focus on only the nuclear flux, and not the host galaxies, for these objects, which is especially important for those at lower redshifts. We applied aperture corrections to these photometric data using empirical JWST/NIRCam PSFs assuming point source morphologies for our sources \citep[see][for more details]{ji2023}. The PSFs used in this analysis were derived from pre-flight pointing predictions. Below, we will discuss the samples of objects from the literature that we explore, and how we matched their positions to sources in our own catalog. We make note of sources that appear in multiple catalogs. Some sources appear in areas with only partial NIRCam coverage, which does not allow us to measure the colors, morphologies, or slopes discussed in the subsequent analysis, so they are removed from our full sample. The data described here may be obtained from the JADES MAST archive at \dataset[doi:10.17909/8tdj-8n28]{https://dx.doi.org/10.17909/8tdj-8n28} \citep{JADES_MAST} and the CEERS MAST archive at \dataset[doi:10.17909/z7p0-8481]{https://dx.doi.org/10.17909/z7p0-8481} \citep{CEERS_MAST}.

\subsection{Little Red Dot Catalogs}

Our primary samples of LRDs were presented in \citet{perezgonzalez2024} for sources in GOODS-S, who used color selection and in \citet{kocevski2024} in GOODS-S and the EGS, where the authors found LRDs based on fits to their photometry. Both methods are described in Section \ref{sec:selection}. The \citet{perezgonzalez2024} sample had 31 sources, and we match these sources in our photometric catalog. In \citet{kocevski2024}, the authors presented 341 LRDs across a number of surveys. We specifically looked at the 120 sources they identified in the JADES and NGDEEP areas in GOODS-S, and in the CEERS survey area in the EGS. We cross-matched these sources to our photometric catalog and recovered 119 sources (the final source, CEERS 19799, is very near a bright galaxy and does not appear in the catalog we matched to). There is an overlap of 17 sources between the \citet{perezgonzalez2024} and \citet{kocevski2024} catalogs in GOODS-S, and we consolidated these sources as belonging to the former catalog due to the date of their initial discovery.

\subsection{Type I AGNs}

To explore how LRD properties relate to AGNs, we also collected samples of spectroscopically-selected Type I AGNs from the literature across GOODS-S, GOODS-N, and CEERS. Our primary GOODS-S and GOODS-N sample of broad-line, Type I AGNs is from \citet{maiolino2023} and Juod{\v{z}}balis et al. (in prep), which were selected from JWST/NIRSpec observations of JADES targets. Most of these AGNs were found as part of the JADES NIRSpec target selection as outlined in \citet{bunker2023} and \citet{deugenio2024}. These were selected based on their photometric redshift from galaxies with normal star forming or quiescent colors. However, a small subset were specifically targeted because they had LRD colors and morphologies, and we highlight those in subsequent figures. We supplement this sample with the $z = 5.55$ broad-line AGN in the system GS-3073 from the GA-NIFS survey \citep{ubler2023,ji2024}, for a total of 17 NIRSpec-selected sources in GOODS-S and 15 sources in GOODS-N. There are four sources from the \citet{perezgonzalez2024} and \citet{kocevski2024} LRD catalogs that are JADES NIRSpec Type I AGNs, and we do not double-count them in the former two catalogs. We indicate spectroscopically-confirmed broad-line sources in subsequent figures using dark outlines. A subsample of the \citet{kocevski2024} LRDs in CEERS have been observed to have broad Balmer emission lines, which we will discuss later in Section \ref{sec:results}. The spectroscopy used to find these lines was observed as part of both the CEERS ERS observations \citep{kocevski2023} and the Red Unknowns: Bright Infrared Extragalactic Survey (RUBIES) \citep{degraaff2024}. 

We also include the GOODS-N and GOODS-S broad-line AGNs found using FRESCO NIRCam grism observations in \citet{matthee2024}. While the authors present 8 sources found across both areas, we only cross-matched to 7, as one source is on a diffraction spike from a nearby star. The only \citet{matthee2024} source in GOODS-S is also found in the JADES NIRSpec Type I sample, so we do not include it here, again to avoid double-counting, although we highlight it in figures. We also include 11 broad H$\alpha$ sources selected from the Complete NIRCam Grism Redshift Survey (CONGRESS) (PID 3577, PID Egami) observations in GOODS-N from Zhang et al. (in prep). This paper also includes 3 additional broad-line sources identified from the FRESCO NIRCam grism data that were not found in \citet{matthee2024}, as these sources were either blended with nearby galaxies, or at lower broad-H$\alpha$ luminosities. This results in a total of 14 additional Type I AGNs, which we cross-match to sources in our photometric catalog. We also include a sample of 12 Type I AGNs in GOODS-S and GOODS-N at $z = 2 - 4$ presented in \citet{sun2024} with broad near-infrared emission lines (Pa $\alpha$, Pa $\beta$, and He\,I $\lambda$10833\,\AA) observed with the NIRCam grism. There is one additional source in their sample at $z > 2$, GS-206907, but this also appears in the JADES NIRSpec Type I sample from Juod{\v{z}}balis et al. (in prep), so we do not include it in the \citeauthor{sun2024} sample. Finally, we add in the GOODS-N broad-line AGN, GN-72127 ($z = 4.13$) described in \citet{kokorev2024b}. 

For the CEERS area, we include the 8 broad-line sources from \citet{harikane2023} (there are two additional sources in this study that are found outside of CEERS in a separate survey). After cross-matching, we only find that five have NIRCam filter coverage and can be used in our study. We remove two sources from the \citet{kocevski2024} CEERS sample as they are also found in the \citet{harikane2023} sample.

\subsection{Type II AGNs and SED-derived AGNs}

We also include a sample of Type II candidate AGNs across GOODS-S described in \citet{scholtz2023} and selected using JADES NIRSpec data. These sources were identified using a variety of methods, including exploring the presence of strong high-ionization emission lines in the spectra for these sources and comparing observed emission line ratios to star formation and AGN models. There are 41 objects in this sample, all of which had counterparts in our catalog. For CEERS, we added the sample of Type II AGNs from \citet{mazzolari2024}, who found 52 objects in a similar manner to what was done in \citet{scholtz2023}. We cross-matched these objects to our catalog and only found 25 sources, and the remaining are off of the NIRCam footprint in the region.

Finally, in order to explore a larger range of redshifts and colors of AGNs with infrared coverage, we also include a sample of AGNs at $z > 2$ from \citet{lyu2024} in GOODS-S. Here, the authors use multi-band JWST/MIRI data at 5--25 $\mu$m from the Systematic Mid-Infrared Instrument (MIRI) Legacy Extragalactic Survey (SMILES) \citep{rieke2024,alberts2024} and JWST/NIRCam and HST data at shorter wavelengths to identify AGNs based on fits to the source optical to mid-IR SEDs. The 124 sources in this sample have both spectroscopic and photometric redshifts, and we adopt their redshifts for the purposes of this study. We cross-matched to 121 sources in our final sample, and removed the 5 sources that are also JADES NIRSpec Type I or Type II AGNs.

\subsection{Sample Summary}

\begin{figure}
  \centering
  \includegraphics[width=0.99\linewidth]{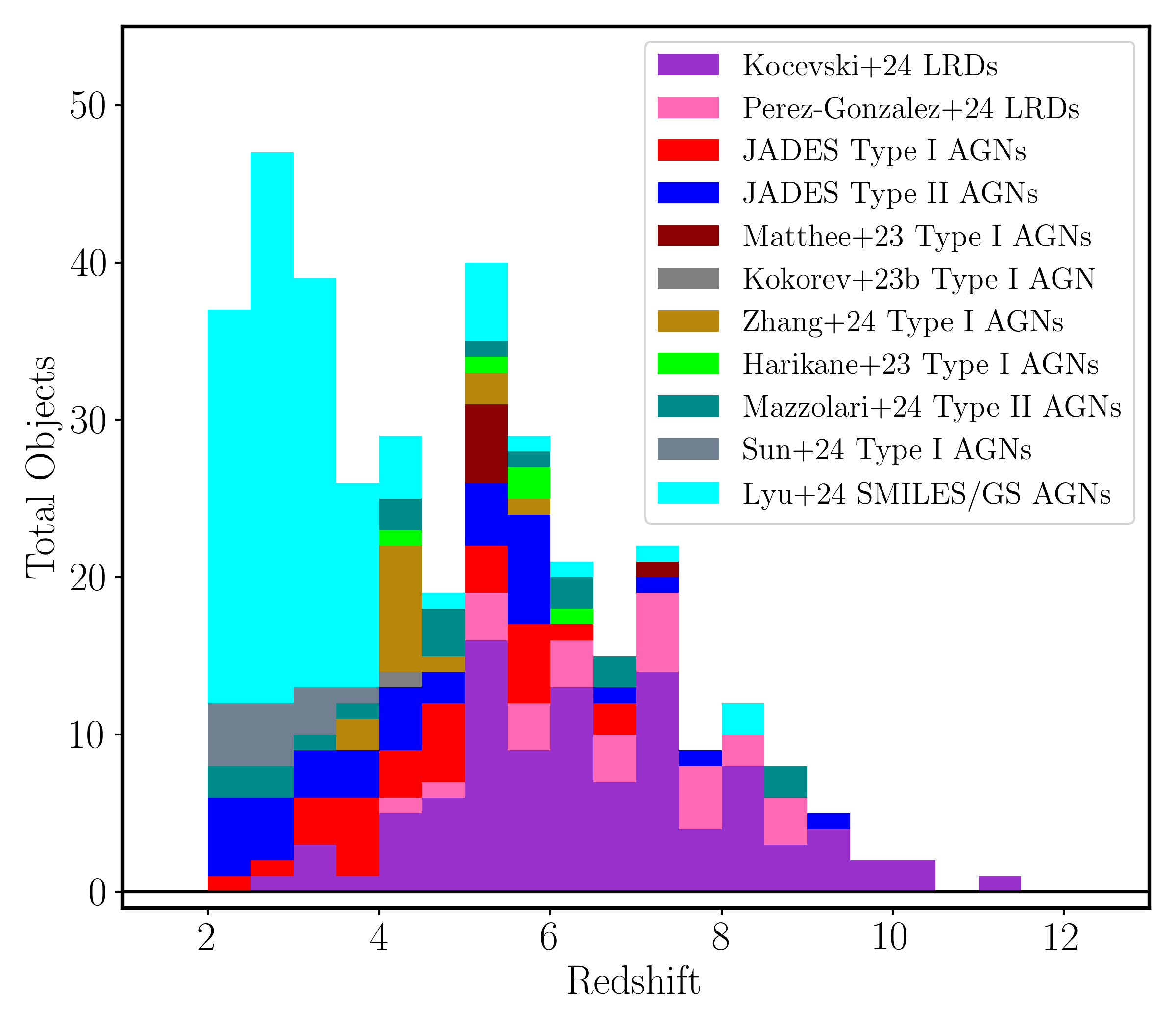}
  \caption{Photometric and spectroscopic redshift stacked histogram for the assembled AGN and LRD samples. Each bar in the stacked histogram shows all of the sources that went into that bin, without any overlapping between the different samples. For the LRD sources, we plot the JADES+NGDEEP+CEERS LRD photometric and spectroscopic redshifts from \citet{kocevski2024} with purple and the JADES LRDs from \citet{perezgonzalez2024} with pink. For the broad-line sources, we plot the spectroscopic redshifts for Type I AGNs in JADES from \citet{maiolino2023} and Juod{\v{z}}balis et al. (in prep) in red, the spectroscopic redshifts for the broad-line AGNs found using FRESCO in \citet{matthee2024} in maroon, the spectroscopic redshifts for the broad-line AGNs found in CEERS in \citet{harikane2023} in lime green, and the Type I AGNs from \citet{sun2024} in slate grey. For obscured sources, we plot the spectroscopic redshifts for Type II AGNs in JADES from \citet{scholtz2023} in dark blue, and the photometric and spectroscopic redshifts for SED-derived AGNs at $z > 2$ across JADES/SMILES in \citet{lyu2024} in light blue. In these plots, objects are only counted once, even if they are contained in more than one sample.}
  \label{fig:redshift_distribution}
\end{figure}

We summarize our final assembled and cross-matched sample and include the number of objects, the minimum and the maximum redshifts for each subsample in Table \ref{tab:lrd_agn_samples}. Additionally, we plot the redshift distribution for the subsamples in Figure \ref{fig:redshift_distribution}, where each bar represents a stack of the individual samples. Here, we show either photometric or spectroscopic redshift, although only a subset of the \citet{kocevski2024} LRDs and \citet{lyu2024} AGNs have spectroscopic redshifts. The \citet{kocevski2024} sample spans the widest range ($z = 3 - 12$), while the Type I and Type II samples only extend between $z = 2 - 8$, with a sole JADES Type II AGN at $z > 8$. The \citet{lyu2024} SED-derived AGNs are primarily found at $z = 2 - 4$ because of the MIRI depth and coverage in the SMILES survey, but we include these sources as a color comparison to samples of more local AGNs. The samples of LRDs we assemble rise at $z > 4$, peak at around $z = 6-8$, and then fall at higher redshifts. \citet{kocevski2024} surmise that the lack of LRDs at $z < 4$ is primarily indicative of a lack of red optical sources with UV excesses at low redshifts, rather than an effect of galaxies being more extended at lower redshifts. The fall-off in the distribution at $z > 8$ is likely due to the rest-frame optical being redshifted out of NIRCam coverage, and to the lower number density of galaxies at all types at progressively higher redshifts.

\begin{deluxetable}{l l r r r}
\tabletypesize{\footnotesize}
\tablecolumns{5}
\tablewidth{0pt}
\tablecaption{Assembled GOODS-S and CEERS LRD and AGN Sample Properties \label{tab:lrd_agn_samples}}
\tablehead{\colhead{Source} & \colhead{Field} & \colhead{$N_{\mathrm{obj}}$} & \colhead{$z_{\mathrm{min}}$} & \colhead{$z_{\mathrm{max}}$}  }
\startdata
	\hline
		\multicolumn{5}{c}{Little Red Dots}\\
	\hline 
		\citet{perezgonzalez2024} & GOODS-S &  27 & 4.1 &  8.8 \\
		\citet{kocevski2024}       & GOODS-S &  35 & 2.9 & 11.3 \\
		                             & CEERS   &  61 & 3.4 & 10.2 \\
	\hline 
		\multicolumn{5}{c}{Type I AGNs}\\
	\hline 
		Juodzbalis et al. (in prep) \&  & GOODS-S &  16 & 3.2 & 6.3 \\
		\phantom{XXX}\citet{maiolino2023}       & GOODS-N &  14 & 1.7 & 6.8 \\
		\citet{ubler2023}          & GOODS-S &   1 &  \multicolumn{2}{c}{5.5} \\
		\citet{matthee2024}        & GOODS-N &   6 & 5.1 & 7.5 \\
		\citet{sun2024}            & GOODS-S &   4 & 2.6 & 3.6 \\
		                             & GOODS-N &   8 & 2.0 & 3.4 \\
		\citet{kokorev2024b}       & GOODS-N &   1 & \multicolumn{2}{c}{4.1} \\
		\citet{harikane2023}       & CEERS   &   5 & 4.5 & 6.0 \\
		Zhang et al. (in prep)          & GOODS-N &  14 & 3.9 & 5.5 \\
	\hline
		\multicolumn{5}{c}{Type II AGNs}\\
	\hline
		\citet{scholtz2023}        & GOODS-S &  41 & 0.7 & 9.4 \\
		\citet{mazzolari2024}     & CEERS   &  18 & 2.4 & 8.9 \\
	\hline
		\multicolumn{5}{c}{SED-derived AGNs}\\
	\hline
		\citet{lyu2024}            & GOODS-S & 114 & 2.0 & 8.4 \\
\enddata
\end{deluxetable}

\section{LRD Selection Methods} \label{sec:selection}

\begin{figure*}
  \centering
  \includegraphics[width=0.99\linewidth]{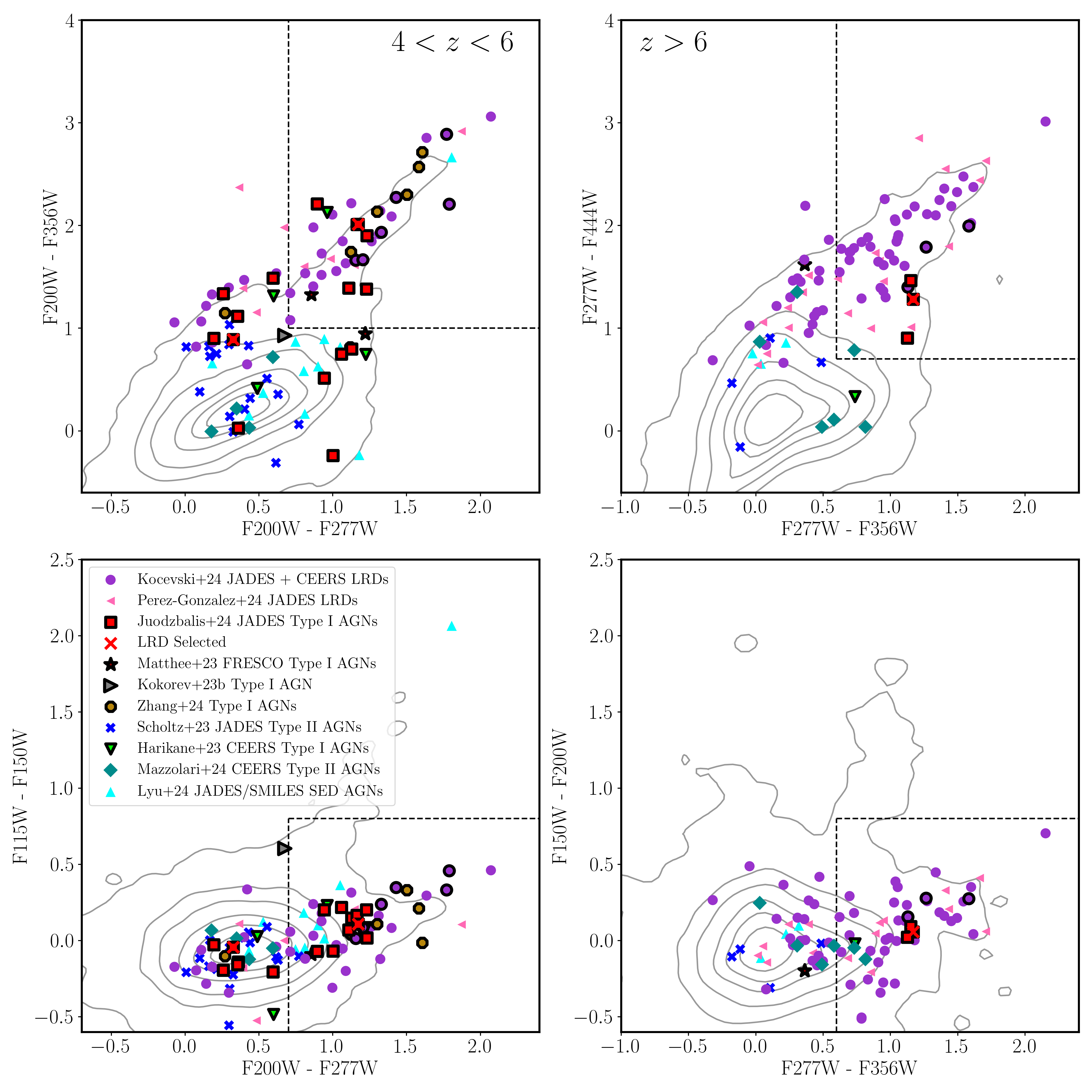}
  \caption{Color-color plots with criteria used to select LRDs from \citet{greene2024}. In the left column we plot color criteria for galaxies at $4 < z < 6$, with the ``red 1'' selection limits as dashed lines. Similarly, in the right column, we plot color criteria for galaxies at $z > 6$, with the ``red 2'' selection limits. The top row shows criteria designed to select for red optical colors, while the bottom row has criteria designed to select blue UV colors. As purple circles we plot the JADES+CEERS LRDs from \citet{kocevski2024}, and as pink left-pointing triangles, the JADES LRDs from \citet{perezgonzalez2024}. As red squares we plot the Type I AGNs in JADES from \citet{maiolino2023},  Juod{\v{z}}balis et al. (in prep), and \citet{ubler2023}. Three of these objects were initially selected for spectroscopic follow-up by virtue of their LRD-like colors, and we highlight those sources with red X's. As maroon stars, we plot the broad-line AGNs from FRESCO/GOODS-N in \citet{matthee2024}. In the lime colored downward-pointing triangles we plot the broad-line AGNs from \citet{harikane2023}. As a grey rightward-pointing triangle we plot the Type I AGN from \citet{kokorev2024b}. For ease of viewing, each spectroscopically-confirmed broad-line source is plotted with a black outline. As dark blue X's we plot the JADES Type II AGNs from \citet{scholtz2023}, as light blue upward-pointing triangles we plot the SED-derived AGNs at $z > 2$ from \citet{lyu2024}, and as turquoise diamonds we plot the Type II AGNs from CEERS assembled in \citet{mazzolari2024}. The contours in each plot show the distribution of JADES+CEERS sources with SNR $> 5$ in each of the filters being used in each panel.}
  \label{fig:color_color_selection}
\end{figure*}

The first samples of photometrically-selected LRDs were chosen using NIRCam colors that probed the rest-frame UV and rest-frame optical portions of the galaxy's SED at a given redshift, along with a cut on compactness that targeted unresolved sources. The LRD color selection in \citet{greene2024} was designed with this in mind, with an additional cut to remove brown dwarfs following work done by \citet{langeroodi2023}, \citet{burgasser2024} and \citet{hainline2024}. Brown dwarfs, especially T and Y dwarfs with $T_{\mathrm{eff}} < 1300K$, have NIRCam colors that rise at 3 - 4 $\mu$m similar to LRDs, but often with extreme blue 1 - 2 $\mu$m colors, which can be selected against. 

The \citet{greene2024} selection colors are, at $4 < z < 6$ (referred to as ``red 1''):

\[
\mathrm{F115W} - \mathrm{F150W} < 0.8 \;\& 
\]
\[
\mathrm{F200W} - \mathrm{F277W} > 0.7 \; \& 
\]
\[
\mathrm{F200W} - \mathrm{F356W} > 1.0
\]

and, at $z > 6$ (referred to as ``red 2''):

\[
\mathrm{F150W} - \mathrm{F200W} < 0.8 \;\& 
\]
\[
\mathrm{F277W} - \mathrm{F356W} > 0.7 \; \& 
\]
\[
\mathrm{F277W} - \mathrm{F444W} > 1.0
\]

We plot these colors for our fully assembled sample in Figure \ref{fig:color_color_selection}, where the left column plots are using the ``red 1'' criteria, and are only those objects at $4 < z < 6$, and the right column plots are using the ``red 2'' criteria with objects at $z > 6$. On these plots, we also show, with grey contours, those objects in the full JADES+CEERS survey with SNR $> 5$ in each of the four filters in each panel, to demonstrate the colors of the bulk of the underlying galaxy population. 

\begin{figure*}[t]
  \centering
  \includegraphics[width=0.48\linewidth]{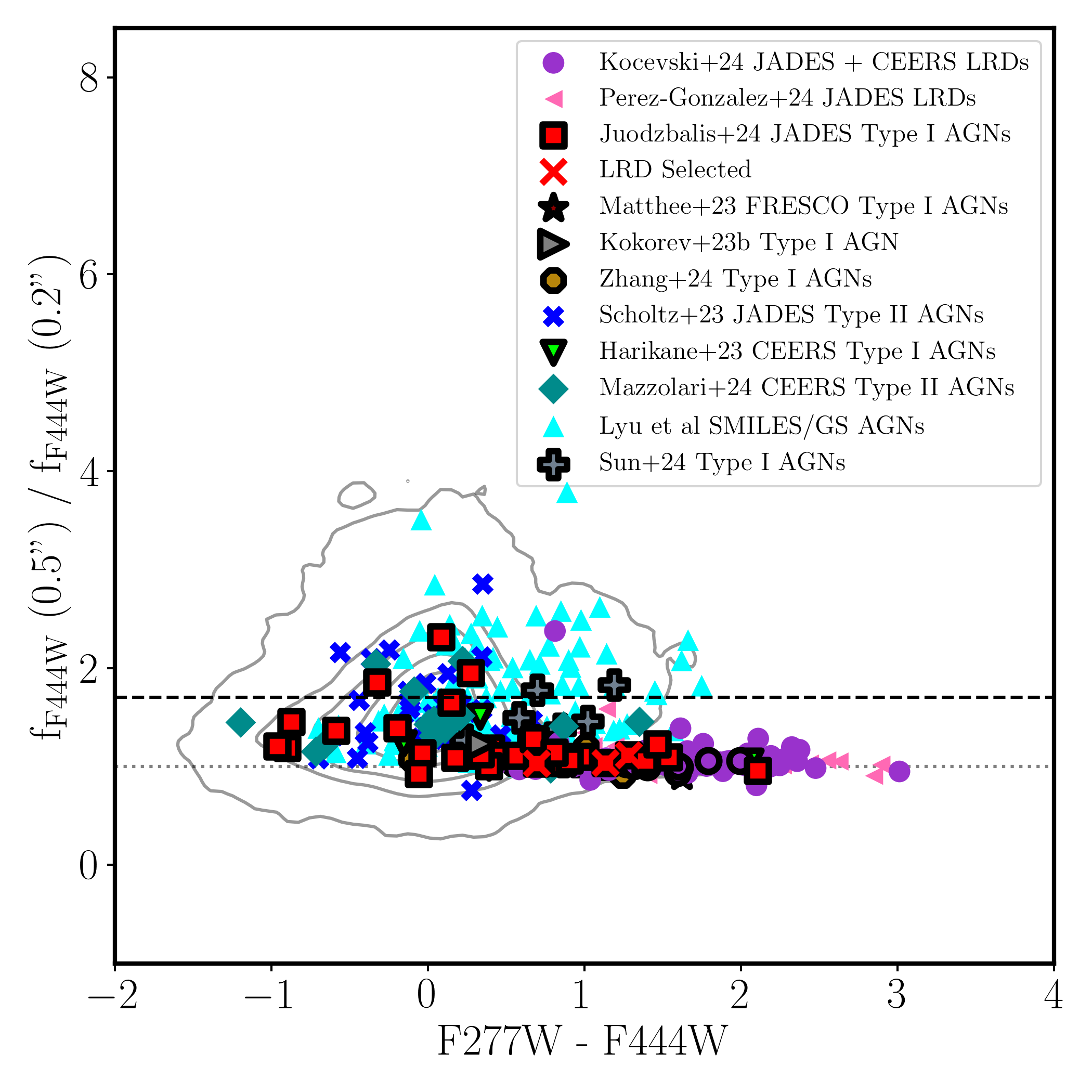}
  \includegraphics[width=0.48\linewidth]{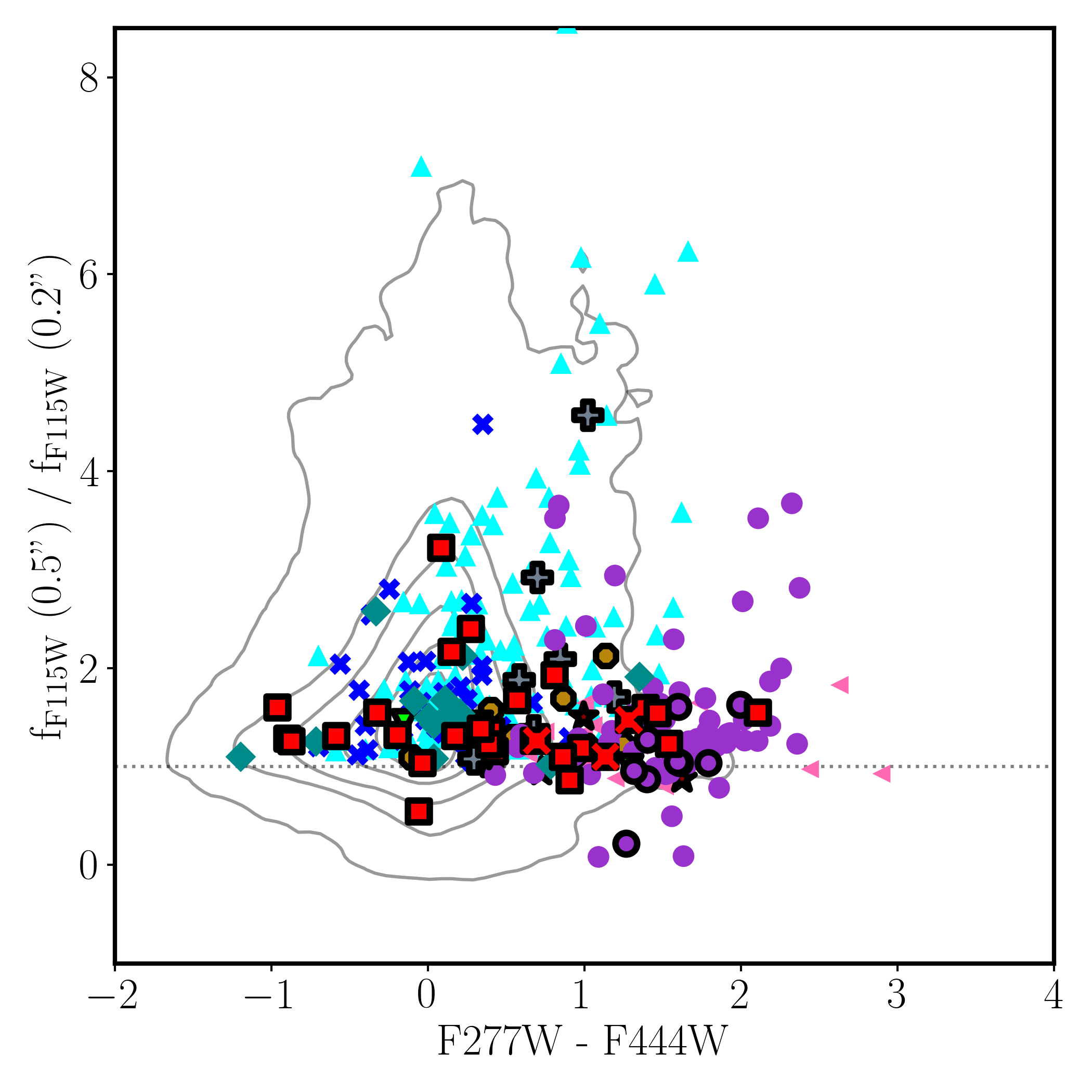}
  \caption{(Left) The F444W compactness criterion is the ratio between the flux in a $0.5^{\prime\prime}$ diameter circular aperture to one in a $0.2^{\prime\prime}$ diameter circular aperture. Here we plot the compactness ratio against the F277W - F444W color for the objects with the same points and colors as in Figure \ref{fig:color_color_selection}, and represent the maximum ratio used for selecting LRDs in \citet{greene2024} with a black dashed line. (Right) The same as the left panel, but for the NIRCam F115W compactness criterion, which probes the rest-frame UV, showing a significantly larger scatter in the sizes for the sources.  The fluxes we use to calculate the compactness criterion in both panels were PSF-corrected assuming a point source, and in both panels we show a ratio of 1.0 with a grey dotted line. }
  \label{fig:compactness_vs_color}
\end{figure*}

Additional criteria were introduced by \citet{greene2024}, who required a significantly bright F444W magnitude, as well as a compactness criterion measured by the ratio of the F444W fluxes in a larger to a smaller circular aperture. We plot the compactness criteria:

\[
f_{\mathrm{F444W}}(0.5^{\prime\prime})/f_{\mathrm{F444W}}(0.2^{\prime\prime}) < 1.7
\]

against the F277W - F444W color for all of the sources in our assembled sample in the left panel of Figure \ref{fig:compactness_vs_color}. We caution that \citet{greene2024} define the compactness criterion as $f_{\mathrm{F444W}}(0.4^{\prime\prime})/f_{\mathrm{F444W}}(0.2^{\prime\prime})$, but, because of the larger range in redshifts that we are probing, we are using a larger aperture. In the right panel of Figure \ref{fig:compactness_vs_color}, we instead plot the compactness criterion but where we use the NIRCam F115W filter, which probes the rest-frame UV for the sources in our assembled sample. Because the rest-frame UV is fainter in a number of these sources, we only plot those objects where the SNR in F115W is greater than 5 in the right panel. In both panels, we plot in contours those sources with SNR $> 5$ in both the F277W and F444W filters in the JADES and CEERS survey, and include a dotted grey line to indicate a ratio of 1.0. While LRDs are often very compact in F444W (for \citet{kocevski2024}, this is part of their selection), there is significantly more scatter in F115W size. Partly this is a result of the LRDs being fainter in the rest-frame UV, and for the most extreme ratios, the LRD being proximate to a lower redshift source that contaminated the $0.4^{\prime\prime}$ aperture. The median F115W compactness criterion value for the \citet{perezgonzalez2024} LRDs in our assembled sample is 1.24 (with a standard deviation of 0.26) and for the \citet{kocevski2024} LRDs the median is 1.27 (with a standard deviation of 1.94), compared to 1.04 (and a standard deviation of 0.15) and 1.03 (and a standard deviation of 0.26) measured using the F444W compactness criterion. The higher fraction of more extended sources indicates that the rest-frame UV may be probing galaxy emission. A more thorough exploration of the rest-frame UV sizes of LRDs will be undertaken in a separate paper \citep{rinaldi2024}.

The use of color cuts is the most straightforward way to select LRDs, but given the large spread in redshifts for galaxies in JWST/NIRCam samples, alternate methods have been developed to select for red optical slopes and blue UV slopes. In \citet{kocevski2024}, the authors fit a series of photometric bands in multiple redshift bins to estimate the UV slope ($\beta_{\mathrm{UV}}$) and optical slope ($\beta_{\mathrm{opt}}$) to select for LRDs. At $z = 2 - 3.25$, $\beta_{\mathrm{UV}}$ is estimated from the HST/ACS F606W, F814W and JWST/NIRCam F115W filters, while $\beta_{\mathrm{opt}}$ is estimated from the NIRCam F150W, F200W, and F277W filters. At $z = 3.25 - 4.75$, $\beta_{\mathrm{UV}}$ is estimated from the ACS F814W and NIRCam F115W and F150W filters, while $\beta_{\mathrm{opt}}$ is estimated from the NIRCam F200W, F277W, and F356W filters. At $z = 4.75 - 8.0$, $\beta_{\mathrm{UV}}$ is estimated from the NIRCam F115W, F150W, and F200W filters, while $\beta_{\mathrm{opt}}$ is estimated from the F277W, F356W, and F444W filters. Finally, at $z > 8$, $\beta_{\mathrm{UV}}$ is estimated from the NIRCam F150W, F200W, and F277W filters, while $\beta_{\mathrm{opt}}$ is estimated from the F356W and F444W filters. 

We estimated $\beta_{\mathrm{UV}}$ and $\beta_{\mathrm{opt}}$ for our full assembled sample of galaxies at $z > 2$ following the method from \citet{kocevski2024}, and we plot the results in Figure \ref{fig:uv_beta_slope_selection}. We note that while the LRDs are chosen to be bright in the rest-frame optical, many sources are faint in the rest-frame UV, which makes accurate estimations of the UV slope more difficult. The LRD selection method in \citet{kocevski2024} allows for a large range in $\beta_{\mathrm{UV}}$ values as a result. 

In \citet{kocevski2024}, the authors translate the typical LRD selection color limits into cuts on $\beta_{\mathrm{UV}}$ and $\beta_{\mathrm{opt}}$: $\beta_{\mathrm{opt}} > 0$, $\beta_{\mathrm{UV}} < -0.37$, and $\beta_{\mathrm{UV}} > -2.8$. The first cut selects for the red optical slope typical of LRDs, the second cut selects for the (relatively) blue UV slope, and the final cut selects against brown dwarfs. We plot these color selection criteria with dashed lines in Figure \ref{fig:uv_beta_slope_selection}. While some of the points from the JADES+NGDEEP+CEERS \citet{kocevski2024} sample fall outside the selection boxes, this is due to the different photometry (the reduction methodology, the usage of non-PSF-convolved mosaics, and different-sized apertures) from what we used in our analysis. Given the dependence of the measurement of $\beta_{\mathrm{UV}}$ and $\beta_{\mathrm{opt}}$ on redshift, the grey contours in this Figure are chosen from a sample of sources in GOODS-S and GOODS-N that have spectroscopic redshifts at $z > 2$, and trace the bright underlying galaxy population. 

As a final comparison of broad-line AGNs, we also include the LRDs from \citet{greene2024}, selected from the Ultra-deep NIRCam and NIRSpec Observations Before the Epoch of Reionization \citep[UNCOVER,][]{bezanson2022} survey. While we do not measure the UV and optical slopes from our own photometry, we use the values provided in Table 2 of \citet{greene2024}.

We additionally plot the location of unobscured AGN slopes in Figure \ref{fig:uv_beta_slope_selection} with vectors that show the effects of dust attenuation. We first plot the UV and optical slopes for the Sloan Digital Sky Survey (SDSS) Type I QSO composite spectrum from \citet{vandenberk2001} with a dark green vertical hexagon. This composite spectrum was derived from a median combination of the spectra for 2,200 SDSS quasars at $z < 4.8$. There is a possibility of contamination in this median spectrum from the host galaxy of these quasars, and additionally, the quasars used in the stack are likely affected by some dust reddening \citep{Richards2003}. Therefore, we also plot with a dark red horizontal hexagon a point representing the characteristic slope for a ``pure accretion (slim) disk'' ($\beta = -2.33$) as derived in \citet{lyndenbell1969}.

We plot two lines representing dust attenuation for each of these models for unobscured quasars. The more horizontal line represents the average dust extinction derived for the Small Magellanic Cloud (SMC) in \citet{gordon2024}. The more vertical line is the extinction law derived from observations of active galactic nuclei in \citet{gaskell2004}. This law, which is relatively flat in the UV potentially due to a lack of small dust grains near the AGN central region, was invoked by \citet{li2024} as a possible origin for the UV and optical slopes observed for LRDs. We note that the derivation of this ``grey'' extinction law may be a result of the inclusion of higher redshift quasars to probe the rest-frame UV, as these sources have less dust extinction \citep{willott2005}. For both laws, we color the points by the value for $\mathrm{A}_V$ as shown in the color bar to the right of the Figure. 

\section{Results} \label{sec:results}

\subsection{AGNs vs. LRDs Spectral Slopes and Extinction Effects}

\begin{figure*}
  \centering
  \includegraphics[width=0.99\linewidth]{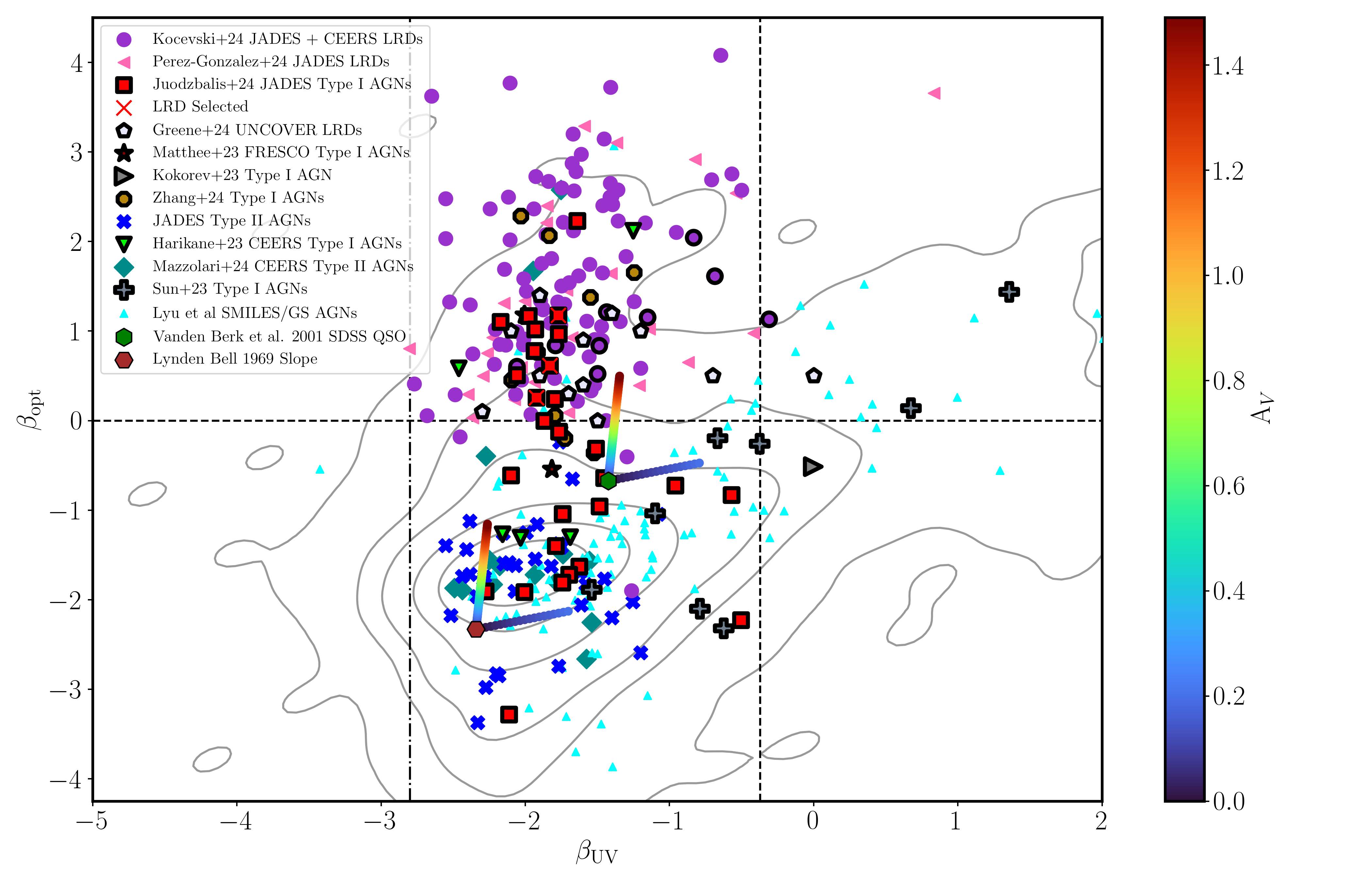}
  \caption{Optical slope ($\beta_{\mathrm{opt}}$) plotted against ultraviolet slope ($\beta_{\mathrm{UV}}$) for the assembled sample. The slopes were measured using the same JWST/NIRCam filter combinations and methods as discussed in \citet{kocevski2024}. We plot the same points as in Figure \ref{fig:color_color_selection}, but we add in the spectroscopically confirmed broad-line UNCOVER LRDs from \citet{greene2024}, and the contours correspond to sources with spectroscopic redshifts from both JADES GOODS-S and GOODS-N. We additionally plot two points representing Type I AGNs: we show the SDSS composite quasar SED from \citet{vandenberk2001} with a green vertical hexagon, and the \citet{lyndenbell1969} characteristic slope for a pure accretion disk ($\beta = -2.33$) with a dark red horizontal hexagon. For both of these points we show how extinction would change the colors of these templates, with the more horizontal bar showing SMC-like dust attenuation from \citet{gordon2024} and the more vertical bar showing the grey extinction from \citet{gaskell2004}. For both lines, the points are colored by $\mathrm{A}_V$ as shown in the color bar on the right hand side of the Figure.
  }
  \label{fig:uv_beta_slope_selection}
\end{figure*}

Together, Figures \ref{fig:color_color_selection}, \ref{fig:compactness_vs_color} and \ref{fig:uv_beta_slope_selection} demonstrate the variety of colors, sizes, and slopes from which LRDs have been selected from samples of galaxies in the literature. It is not surprising that color selection criteria from \citet{greene2024} and \citet{kokorev2024} do not select for all of the sources found using the fit UV and optical slope criteria from \citet{kocevski2024}, as seen in Figure \ref{fig:color_color_selection}, because of the impact of redshift on the observed-frame SED. The same figures highlight that a large fraction of AGN identified spectroscopically by JWST (without pre-selecting by color or size), do not meet any of the LRD selection criteria. Indeed, LRDs occupy an extreme color space when compared to the underlying galaxy population and also when compared to other samples of both Type I and Type II AGNs.

We find that both the \citet{scholtz2023} and \citet{Mazzolari24_CEERS} (Type II selected) and \citet{lyu2024} (mid-IR selected) AGN populations have colors similar to the underlying galaxy population and we observe extended morphologies for these sources. This is not surprising as, due to nuclear obscuration, these sources have UV and optical colors dominated by the host galaxy. While most of the Type II AGNs are found at $z < 6$, even for those at $z > 6$, only two spectroscopically-confirmed Type II AGNs are found with optical and UV slopes that would be classified as LRDs in Figure \ref{fig:uv_beta_slope_selection}. These two sources are located at $z \sim 8.7$, and at this redshift [\ion{O}{3}]$\lambda\lambda4959,5007$+H$\beta$ emission may be boosting the F444W filter, causing an observed red slope. The lack of Type II AGNs with redder values of $\beta_{\mathrm{opt}}$ may suggest that the red optical slope observed in LRDs may be associated with the accretion disk that is otherwise completely obscured in the Type II sources.

We can see that the underlying spectroscopic-redshift sample of JADES has a bimodality in Figure \ref{fig:uv_beta_slope_selection}, where the majority fall along a sequence towards red UV and optical colors (the top right of the figure), while a smaller subset move upward to red optical slopes at a fairly blue UV slope, consistent with LRD colors. Attenuation with an SMC-like extinction law would serve to push galaxies to primarily redder UV slopes, with limited effect on the optical extinction. This direction of attenuation is in agreement with the lower-redshift \citet{lyu2024} SED AGNs, the $z < 4$ broad-line sources from Zhang et al. (in prep), and the spectroscopic galaxy sample shown in the contours. The \citet{gaskell2004} extinction curve, as discussed in \citet{li2024}, would serve to push some of the Type I sources to red $\beta_{\mathrm{opt}}$ values while keeping the $\beta_{\mathrm{UV}}$ relatively similar, but this would require 3 or more magnitudes of visual extinction for many of the sources to reach the red optical slopes observed in LRDs.

The samples of Type I AGNs provide a vital point of comparison. While, as has been shown by numerous authors, many of the broad-line sources in our assembled sample have LRD colors, there are many Type I AGNs that exist outside of LRD selection criteria. In Figure \ref{fig:color_color_selection} this could be due to differences in redshift. As Figure \ref{fig:uv_beta_slope_selection} shows rest-frame slopes, it can be seen that a large number of broad-line sources have relatively blue optical colors, in agreement with the SDSS QSO composite from \citet{vandenberk2001} and even approaching the colors of (unreddened) pure accretion discs \citep{lyndenbell1969}. However, they are also broadly in agreement with typical blue slopes of star-forming galaxies. These properties indicate that the majority of the Type I AGN newly identified by JWST are either normal, blue (accretion-disc-dominated) unobscured (or mildly reddened) AGN, or weak AGN whose continuum is dominated by the host galaxy. The finding that morphologically they are a mixture of extremely compact and extended sources (Fig. \ref{fig:compactness_vs_color}) supports that both cases are actually found.

We note that the broad-line ``quenched'' AGN described in \citet{kokorev2024b} does not fall into any of the LRD color or slope selection criteria based on our photometry, despite the claims by those authors that this source is an LRD. From the circular photometry we use in this paper, it is near the edge of the color selection criteria in Figure \ref{fig:color_color_selection}, but the UV slope is too red, and the optical slope is significantly too blue, to be selected by the \citet{kocevski2024} slope criteria as shown in Figure \ref{fig:uv_beta_slope_selection}.

\begin{figure}
  \centering
  \includegraphics[width=0.99\linewidth]{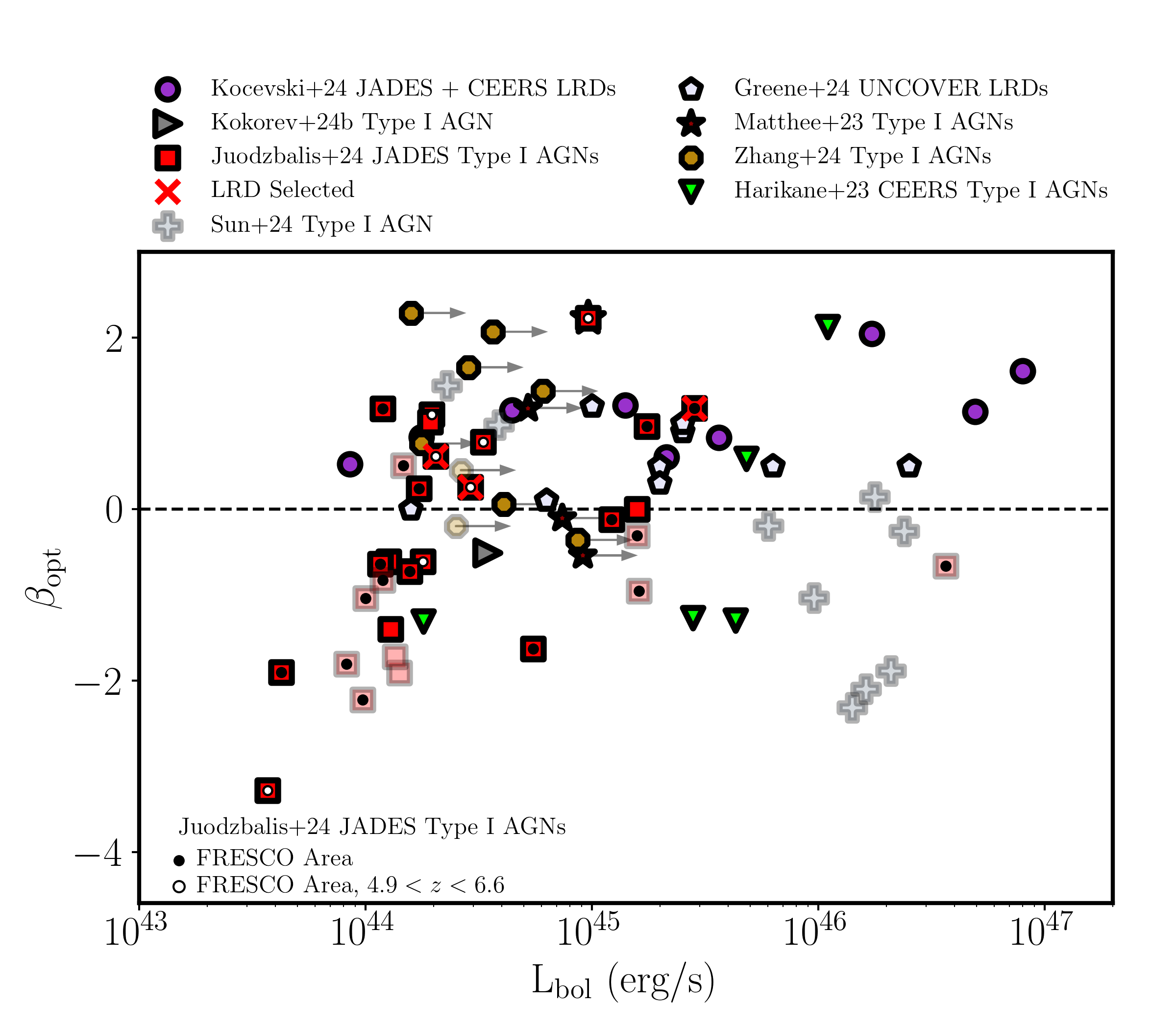}
  \caption{Rest-frame optical slope plotted against AGN bolometric luminosity for the broad-line sources in our assembled samples. The points are the same as in Figure \ref{fig:uv_beta_slope_selection}. Bolometric luminosities were estimated from fits to the broad H$\alpha$, or, in a few cases H$\beta$ emission line, as described in the text. Those sources from multiple samples at $z < 4$ are plotted with lighter-colored points. For the NIRCam grism sources from \citet{matthee2024} and Zhang et al. (in prep), no correction for dust extinction was made to the H$\alpha$-derived bolometric luminosities, so these represent lower limits, represented by the grey arrows. We also indicate those Type I sources from Juod{\v{z}}balis in prep. that sit in the FRESCO area with a dark central circle, and if they are in the FRESCO area and at a redshift of $4.9 < z < 6.6$ with a white central circle. Only one Type I AGN from Juod{\v{z}}balis in prep., in GOODS-S, was also a FRESCO source from \citet{matthee2024}, which we indicate by a square on top of a star. 
  }
  \label{fig:lbol_vs_opt_slope}
\end{figure}

\subsection{Optical Spectral Slopes and AGN Luminosity}

To explore the origin of these differences in optical slope, we compared the optical slope to the AGN bolometric luminosities for the broad-line sources in Figure \ref{fig:lbol_vs_opt_slope}. In this Figure, we used the bolometric luminosities provided by the authors from which the samples were taken. The bolometric luminosities for the Type I AGNs from Juod{\v{z}}balis in prep, \citet{greene2024} and \citet{kokorev2024b} were corrected for dust extinction. We include the \citet{kocevski2024} sources in CEERS with broad-line detections from both \citet{kocevski2023} and from RUBIES observations \citep{degraaff2024}. \citet{kocevski2024} do not provide bolometric luminosities in their study, but they do provide dust-corrected black hole masses and broad H$\alpha$ FWHM values, from which we can derive H$\alpha$ luminosities following Equation 1 in \citet{reines2015}. We then convert these H$\alpha$ luminosities to bolometric luminosities by multiplying them by 130, following \citet{stern2012}. We caution that the conversion between broad-line H$\alpha$ fluxes and bolometric luminosities was derived using AGNs at $z < 0.055$, and these relations may not hold at high redshift, or for LRDs, which may have different densities, distributions, and broad-line region morphologies, as well as intrinsically different SEDs \citep{Maiolino24_Xrays,Pacucci_Narayan_2024,Arcodia2024,Lupi2024}.

For the Type I sources from \citet{sun2024}, we used the bolometric luminosities that they presented derived from SED fits for these sources as described in \citet{lyu2024}. For one source from their sample, GN-1000721, the authors estimated the bolometric luminosity instead from the 2-10 keV X-ray flux and converted this to a bolometric luminosity. For the broad-H$\alpha$ sources found using NIRCam grism spectroscopy in \citet{matthee2024} and Zhang et al. (in prep), the authors do not correct for dust extinction, hence these values should be considered lower limits. In this Figure, we plot those sources in all samples at $2 < z < 4$ with lighter colored points.

We see that the majority (66\% including the lower limits, 62\% excluding the lower limits from the grism surveys) of the broad-line LRDs plotted in Figure \ref{fig:lbol_vs_opt_slope} across multiple samples have $L_{\mathrm{bol}} > 2\times 10^{44}$ erg s$^{-1}$, while 18\% of those sources (21\% excluding the lower limits) with $\beta_{\mathrm{opt}} < 0$ are found at $L_{\mathrm{bol}} < 2\times 10^{44}$ erg s$^{-1}$. There is a hint of a trend at the lowest luminosities for AGN to become bluer and this may be a consequence of the star-forming host galaxy becoming more dominant in such weak AGN.

There is significant scatter in this plot, such that there are also a small sample of luminous broad-line AGNs that do not have red optical colors consistent with LRDs (19 sources have $L_{\mathrm{bol}} > 2 \times 10^{44}$ erg s$^{-1}$ and $\beta_{\mathrm{opt}} < 0$, 15 if you exclude the lower limits), including one source, a Type I AGN at $z = 3.7$ with $L_{\mathrm{bol}} > 10^{46}$ erg s$^{-1}$. The majority of these sources are at $2 < z < 4$, and represent more traditional AGNs.

It is important to note that the broad line sources found as part of blind NIRCam grism surveys, such as the ones in \citet{matthee2024} or Zhang et al. (in prep) do not suffer the same target pre-selection aspects as NIRSpec surveys, although they are limited to luminous sources. The bolometric luminosities for these sources should be considered lower limits as they do not account for any dust reddening to the line. Nonetheless, while there are limited numbers still, the majority (71\%) of the sources from these combined grism spectra surveys have $\beta_{\mathrm{opt}} > 0$. We discuss the possible origin of this fraction in the next section.

\section{Discussion} \label{sec:discussion}

Taken together, these results demonstrate the broad variety of AGNs found at $z > 2$. Figures \ref{fig:color_color_selection}, \ref{fig:uv_beta_slope_selection}, and \ref{fig:lbol_vs_opt_slope} show that, while LRD samples contain a significant fraction of broad-line sources, they may represent a significantly smaller fraction of the total number of broad-line AGNs, as suggested by \citet{harikane2023, labbe2023b, greene2024, maiolino2023, kocevski2024, Taylor2024}. Here, we will discuss the discovery of broad emission lines in Type I AGNs in Section \ref{sec:findingbroadlines}, explore the variety of SEDs found using LRD selection techniques in Section \ref{sec:LRDSelection_and_SEDs}, and estimate the fraction of LRDs among Type I AGNs in our sample in Section \ref{sec:lrdfraction}.

\subsection{Finding Type I AGNs in Samples of LRDs} \label{sec:findingbroadlines}

Multiple effects contribute to the observation of broad Balmer emission lines in LRDs. First, detecting the presence of a broad emission line is easier for those AGNs with higher bolometric luminosities, especially with respect to the underlying narrow H$\alpha$ emission and continuum emission from the host galaxy, and especially at low/intermediate spectral resolution. However, one of the more important issues with understanding the full Type I AGN population, and the role of LRDs, lies in how selecting broad-line AGN candidates for spectroscopic follow-up is difficult for those galaxies where the observed colors match those of typical galaxies. LRDs are a far more straightforward group of sources to select for, and caution must be taken when making broad claims about these sources given that they seem to represent a very luminous population. The majority of the AGNs in these high-redshift galaxies have not been discovered yet, primarily because of a lack of observed spectra for these sources. The only Type I AGN with $\beta_{\mathrm{opt}} < 0$ at $z > 6$, the Type I AGN JADES-GN+189.09147+62.22811 at $z = 6.68$ \citep{juodzbalis2024a}, is on the edge of the selection box in Figure \ref{fig:uv_beta_slope_selection}. Every other broad-line AGN at $z > 6$ has colors consistent with an LRD, demonstrating the current difficulty in selecting broad-line sources at these high redshifts except through targeting these rare objects. % JADES-GN+189.09147+62.22811 is 1001830

One way of mitigating source selection effects is by looking at the samples derived from NIRCam grism spectroscopy. It is difficult to say, however, what the intrinsic luminosity range being probed by the NIRCam grism-derived samples is, because these source luminosities have not been corrected for dust extinction. As shown with black and white dots in Figure \ref{fig:lbol_vs_opt_slope}, a number of the JADES Type I AGNs from Juod{\v{z}}balis et al. (in prep.) are found in the FRESCO survey area in GOODS-S and GOODS-N. The white dots in the Figure indicate which sources within the FRESCO footprint are also found in the redshift range $4.9 < z < 6.6$ where H$\alpha$ would be redshifted into the F444W filter. These sources have dust-corrected $L_{\mathrm{bol}} < 4\times 10^{44}$ erg s$^{-1}$, and broad lines were not detected in the grism spectra, indicating that the FRESCO spectroscopy was not sensitive to lines in this luminosity range. The only Juod{\v{z}}balis et al. source that overlaps with the \citet{matthee2024} sample is JADES-GS+53.1386-27.79025 (GOODS-S-13971, $z = 5.481$), an LRD where the NIRSpec-derived, dust-corrected value is $L_{\mathrm{bol}} = 9.6\times 10^{44}$ erg s$^{-1}$ (the dust-uncorrected value measured in \citeauthor{matthee2024} is $L_{\mathrm{bol}} = 5.5\times 10^{44}$ erg s$^{-1}$). 

There may be also a significant bias to the selection of broad-line AGNs with the NIRCam grism, which is very sensitive to both the flux of a given emission line to be detected (in other words, the wide-band brightness assuming a fixed emission line equivalent width), as well as the morphology of the source \citep[see][for more discussion]{maiolino2023}. Photometric LRD selection, as described in Section \ref{sec:observations}, often includes an upper limit on the F444W magnitude (or a lower limit on the F444W flux). For FRESCO, the F444W filter was used with the NIRCam grism, so it is perhaps not surprising that the sources with broad H$\alpha$ fluxes were red in that sample. In addition, the observed NIRCam grism line width is a convolution of the intrinsic line width, the NIRCam instrumental line broadening, and the source morphology. While there is an observed morphological broadening of grism spectral lines for extended sources, faint broad H$\alpha$ emission has a higher SNR for compact sources. 

\subsection{LRDs Selection, Spectral Energy Distributions, and the Impact of Emission Lines} \label{sec:LRDSelection_and_SEDs}

While a subset of LRDs do represent a distinct sample of Type I AGNs, it is difficult to know exactly whether their large numbers are indicative of trends in AGN evolution at high redshift. Currently, there are many definitions of an LRD in the literature, with multiple selection criteria and techniques. Importantly, these methods rely on colors or slopes estimated from the most commonly-used NIRCam filters, which are relatively wide. Strong line emission in these galaxies, most significantly [\ion{O}{3}]$\lambda\lambda4959,5007$ + H$\beta$ and H$\alpha$, can help boost the fluxes, making a galaxy appear significantly redder at 2 - 5 $\mu$m. Multiple studies have attempted to mitigate this by using NIRCam F410M photometry where it is available to trace the underlying continuum \citep[see][for a discussion]{kocevski2024}, but this is an imperfect technique, given how strong lines can boost both F410M and F444W in specific redshift ranges. Emission lines with high equivalent widths are quite common at $z > 4$ and can have a significant impact on the inferred galaxy properties of red galaxies like LRDs \citep{endsley2023}.

\begin{figure}
  \centering
  \includegraphics[width=1.0\linewidth]{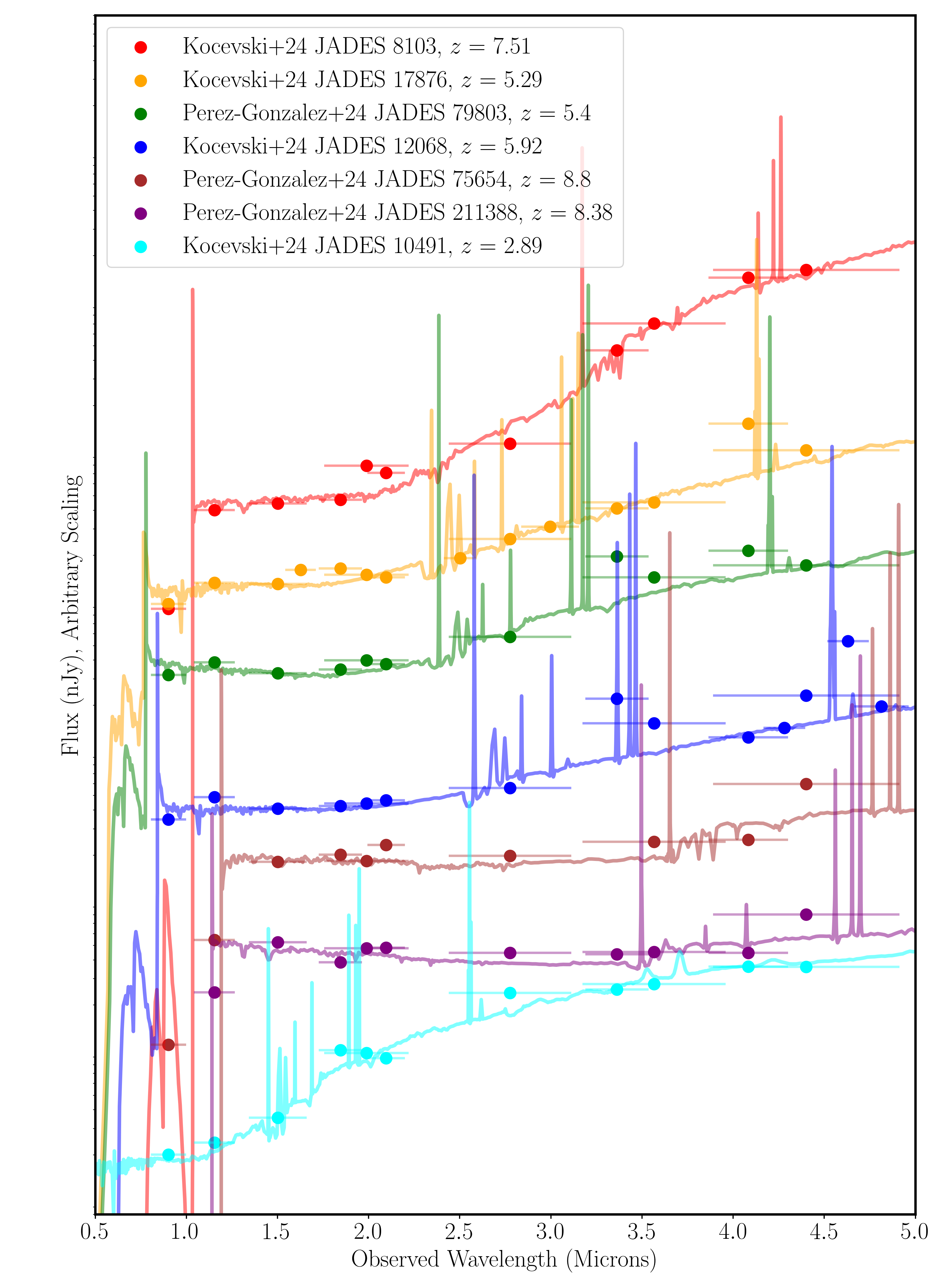}
  \caption{Example NIRCam SEDs for a range of LRDs presented in \citet{kocevski2024} and \citet{perezgonzalez2024}, along with fits to the sources using the python implementation of EAZY \citet{brammer2008}. From top to bottom, we plot the SED for JADES 8103 ($z_{\mathrm{phot}} = 7.51$) and 17876 ($z_{\mathrm{phot}} = 5.29$) from \citeauthor{kocevski2024}, which are examples of LRDs with flat, or blue UV colors and a red optical slope. Below that are \citeauthor{perezgonzalez2024} JADES 79803 ($z_{\mathrm{phot}} = 5.4$) and \citeauthor{kocevski2024} JADES 12068 ($z_{\mathrm{phot}} = 5.92$), where strong optical line emission is boosting the wide filters, and without the medium band filters these sources would be fit with a much redder optical slope. Below that we plot JADES 75654 ($z_{\mathrm{phot}} = 8.8$) and JADES 211388 ($z_{\mathrm{phot}} = 8.38$) from
  \citeauthor{perezgonzalez2024}, which have flat UV slopes and only rise at 4 $\mu$m, likely because of strong [\ion{O}{3}]+H$\beta$ emission. Finally, we plot \citeauthor{kocevski2024} 10491 ($z_{\mathrm{phot}} = 2.89$) a source with a more traditional optical turnover similar to other low-redshift AGNs.
  }
  \label{fig:example_seds}
\end{figure}

To further explore this, we carefully visually inspected each of the photometric SEDs for the Type I AGNs and the LRDs in our assembled samples. As we have shown in Figures \ref{fig:color_color_selection} and \ref{fig:uv_beta_slope_selection}, LRD selection results in samples of objects with a variety of colors and slopes. While the term ``LRD'' could merely refer to compact sources that appear red in the rest-frame optical, there are a number of ways that an observed SED can feature such a slope. To demonstrate this, we plot the photometric SEDs for seven sources from the \citet{kocevski2024} and \citet{perezgonzalez2024} JADES GOODS-S LRD samples in Figure \ref{fig:example_seds}. We plot the observed photometry along with fits to the photometry from the python implementation of \texttt{EAZY}\footnote{https://github.com/gbrammer/eazy-py} \citep{brammer2008}. We use the \texttt{EAZY} templates and fitting procedure outlined in \citet{Hainline2024b}, and fix the redshifts to the values provided by either \citet{kocevski2024} or \citet{perezgonzalez2024}. 

The majority ($\gtrsim 65$\%) of the LRDs in our assembled sample has a rising optical slope, and the NIRCam photometry, specifically the medium band filters found across the JADES fields, provides evidence of this being largely continuum driven. Examples of these sources are shown in the top two SEDs in Figure \ref{fig:example_seds} from the \citet{kocevski2024} sample of LRDs\footnote{both of these sources appear in \citet{perezgonzalez2024}, the former as 203749 ($z_{\mathrm{phot}} = 4.1153$), and the latter as 184838 ($z_{\mathrm{phot}} = 7.1$). Based on our \texttt{EAZY} fits to the SEDs, we consider these redshifts to be less accurate than the values presented in \citet{kocevski2024}, which are shown in Figure \ref{fig:example_seds}.}. While there is emission line boosting from H$\alpha$ in F410M and F444W for \citeauthor{kocevski2024} JADES 17876, this source lies in the JADES Origins Field \citep[JWST PID 3215,][]{eisenstein2023b}, where the shallow red underlying continuum is observed using data from the medium-band filter coverage in this survey area.  

Line boosting is more of an issue for the third and fourth SEDs from the top in Figure \ref{fig:example_seds}, JADES 79803 from \citet{perezgonzalez2024} and JADES 12068 from \citet{kocevski2024}. At the redshifts of these sources, strong [\ion{O}{3}]$\lambda\lambda4959,5007$ line emission is contributing to the F335M and F356W fluxes, and H$\alpha$ is contributing to the F444W fluxes, such that the true continuum slope is difficult to discern from the NIRCam wide filters alone. This is clear from \citeauthor{kocevski2024} JADES 12068, where the NIRCam F410M, F430M, and F480M trace a significantly bluer optical slope than would be estimated from the F356W and F444W filters. For those sources in CEERS, where the only medium-band coverage comes from the F410M filter, this may be a more significant concern, especially for galaxy or AGN properties derived from SED fitting codes. Sufficient medium-band coverage around $3-5$ $\mu$m can isolate line boosting from prominent emission lines and provide more robust estimates of the rest-optical slopes. These SEDs and the fits shown in Figure \ref{fig:example_seds} demonstrate the significant degeneracies that exist between line and continuum emission which biases the selection of LRDs from photometry alone. 

At the highest redshifts ($z > 8$), the rest-frame optical is only probed by data at $\lambda > 4$ $\mu$m, which means that any potential boosting from the [\ion{O}{2}]$\lambda$3727,3729, or [\ion{O}{3}]$\lambda$5007 + H$\beta$ emission lines will result in a red F277W - F444W color, and little information can be gleaned about the underlying optical continuum. This is shown with the fifth and sixth SEDs from the top in Figure \ref{fig:example_seds}, JADES 75654 aand 211388 from \citet{perezgonzalez2024}. It should be noted that the latter was detected with JWST/MIRI at 5.6 - 12.8 $\mu$m indicating a rising optical slope, but this is not guaranteed when selecting LRDs at high redshifts with a red F277W - F444W color. The two Type II AGNs with ``red optical slopes'' seen in Figure \ref{fig:uv_beta_slope_selection} from the \citet{mazzolari2024} sample are only red by virtue of strong line emission in F444W. Similar results are seen for a sample of $z > 7$ galaxies with F277W - F444W $> 1$ in \citet{desprez2024}, where JWST/NIRSpec observations demonstrate that high-equivalent width \ion{O}{3}]$\lambda$5007 + H$\beta$ emission lines contributed $40 - 50\%$ of the observed F444W flux. MIRI observations and medium-band NIRCam observations are potentially necessary to provide more robust estimates of the rest-optical slopes, due to uncertain line contamination in F444W. Additionally, for sources at these redshifts, accurate photometric redshift estimates require deep observations at 0.9 - 1.5 $\mu$m to trace the Lyman-$\alpha$ break. As many of the LRDs are faint in the rest-frame UV, photometric redshift estimates that depend on relatively shallower HST short-wavelength coverage may result in redshifts that are biased high.

At low redshifts, however, LRD selection may be returning sources that are quite different from what is seen at $z > 6$, as shown in the the bottom source in Figure \ref{fig:example_seds}, JADES 10491 from \citet{kocevski2024}. This source has a red UV slope and while the SED does rise in the rest-frame optical, it turns over, and is largely consistent with other low-redshift AGN SEDs.

Taken together, this demonstrates a core issue in understanding the origin of the UV and optical fluxes in LRDs: current selection techniques are too broad, and even in sources with red optical slopes, flux boosting from emission lines plays a significant role. In addition, photometric redshift estimation is difficult for these sources, and in surveys without deep NIRCam observations at $0.9 - 1.5$ $\mu$m, the resulting photometric redshifts are biased high. It is of fundamental importance that careful visual inspection of the SEDs of recovered LRDs is performed so as to specifically target those sources with strong evidence of a V-shaped continuum, with a flat UV slope and a red optical slope. It is crucial that these sources are observed with multiple medium-band filters, given the contributions from strong emission lines. 

\subsection{The Fraction of LRDs Among Type I AGNs}\label{sec:lrdfraction}

As discussed, while the fraction of AGN among LRDs is argued to be between 20\% and 80\%, likely depending on the specific LRD selection criteria \citep{greene2024, perezgonzalez2024, kocevski2024}, it is of interest to explore the fraction of LRDs among the population of Type I, broad-line, AGNs. The heterogeneous sample of Type I AGN collected in this work is certainly inadequate to fully explore this, especially given that many Type I AGNs were discovered specifically in spectroscopic follow-up of LRDs. A potential way forward, although still imperfect, is to use the statistics from the Type I AGNs serendipitously identified from the broad spectroscopic survey of galaxies in JADES (Juodzbalis et al., in prep.).

The selection function of the galaxies targeted in JADES is complex. However, with the exception of a few rare cases targeted specifically for their peculiar colors, the majority of the galaxies targeted in the JADES spectroscopic follow-up are representative of the galaxy population (with magnitude limits varying for different tiers), where the priority was generally given to higher-redshift sources and otherwise the slit allocation was generally provided by the MSA constraints.

Excluding three objects that were specifically targeted for follow-up by virtue of their LRD colors, out of the 27 Type I AGNs identified by Juodzbalis et al. in JADES (down to a bolometric luminosity of a few times $10^{43}$ erg s$^{-1}$), 8 have slopes consistent with an LRD, i.e. about 30\%. However, as already mentioned, there is a strong luminosity dependence. Specifically, at bolometric luminosities below $1.5\times 10^{44}$ erg s$^{-1}$ the fraction is 15\%, and no LRDs are found at $L_{\mathrm{bol}}<10^{44}$ erg s$^{-1}$. As discussed, there is also a possible redshift dependence, however it is more difficult to assess due to the lack of statistics at $z>6$ \citep[and also the fact that the JADES team selected fewer galaxies at low redshifts, see][]{bunker2023, deugenio2024}.

The fraction of LRDs is also dependent on the AGN properties that are considered. For instance, when looking at the UV luminosity function of Type I AGNs (and of their host galaxies), which is relevant for their contribution to the reionization of the Universe \citep{Madau2024,Grazian2024}, the fraction of LRDs is obviously lower owing to their redder observed colors. Exploring the UV luminosity function (and more broadly any luminosity function) of Type I AGN is beyond the scope of this paper. Here we only report that the contribution of LRDs to the UV luminosity function (in the luminosity range between $M_{UV}\sim -18$ and $M_{UV}\sim -20$) has been estimated to be between 3\% and 10\% \citep[e.g.][]{maiolino2023,kocevski2024,Taylor2024}.

\section{Conclusions} \label{sec:conclusions}

To conclude, we have assembled a large sample of LRDs and AGNs from the literature across the JADES and CEERS extragalactic datasets. Using a uniform NIRCam photometric dataset we have explored common color and slope selection criteria for the sources in the sample to understand the role of LRDs in the evolution of AGNs. Primarily, we find that:

\begin{itemize}
  \item Using our independently derived photometric catalogs, we find consistent colors, slopes, and compactness values for JADES and CEERS LRDs to those in the literature, primarily \citet{kocevski2024}. The usage of a rest-frame $\beta_{\mathrm{UV}}$ and $\beta_{\mathrm{opt}}$ for selecting LRDs is a more robust technique than simple color selection for finding sources across a range of redshifts.
  \item The Type I AGNs span a broad distribution of colors and slopes, some of them consistent with LRDs, while others having colors typical of star-forming galaxies or local (blue) quasars (which have similar colors). We find a bimodality in the distribution of UV and optical slopes for both galaxies and AGNs. SMC-like extinction laws would redden both the UV and the optical slopes of Type I sources, producing slopes consistent with intermediate-redshift SED-derived AGNs, as well as some reddened, low-z quasars. The rest-frame UV in LRDs may be dominated by host-galaxy light \citep{killi2023} or scattered light from the accretion disk \citep{kocevski2023, barro2024, labbe2023b, akins2023}, although more grey extinction laws, such as the one derived for AGNs in \citet{gaskell2004}, would serve to move sources into LRD selection regions, as discussed by \citet{li2024}.
  \item The Type II AGNs in our sample do not have colors or slopes consistent with LRDs, but rather they appear to have colors and slopes similar to the underlying galaxy population. This is expected given that central obscuration in these sources results in a continuum dominated by emission from the host galaxy. 
  \item For Type I sources, we find that LRDs typically have high bolometric AGN luminosities compared to the non-LRD Type I AGNs. Because non-LRDs have colors typical of the underlying galaxy population (while being consistent with quasar-like colors, which are similar to those of star-forming galaxies), targeting ``normal'' (blue) Type I AGNs with NIRSpec is not as straightforward as what has been done with LRDs (which have peculiar colors and slopes). 
  \item While NIRCam grism observations can help mitigate this selection effect somewhat, because of how LRDs are selected by being bright at long wavelengths and compact, broad-line sources found in grism data are more likely to arise from an LRD. 
  \item From a careful round of visual inspection, we find that while the majority ($\gtrsim 65$\%) of literature LRD sources have a V-shaped SED, there is significant contamination from emission line fluxes to the photometry used to select these sources, which is more common at higher redshifts ($z > 8$). A full understanding of the role that LRDs play in AGN evolution will require refinements to the selection processes to mitigate this contamination. Deep JWST/MIRI observations can be used to extend the redshift range where LRDs are found.
  \item The majority of Type I AGNs, in the intermediate/low luminosity range newly probed by JWST at $z > 4$, are not LRDs. We attempt to assess the fraction of LRDs among the Type I AGN at high redshift. Taking the JADES spectroscopic survey as a potentially unbiased reference, we find that LRDs are about 30\% of the Type I AGN population, with their fraction dropping to less than 15\% at low AGN bolometric luminosities. Obviously, given their red colors, the contribution of LRDs to the UV luminosity function is even lower, with previous studies estimating fractions between 3\% and 10\%.
\end{itemize}

LRDs remain a fascinating sample of sources, and the origin of their UV and optical emission remains a mystery given what has been observed both spectroscopically and at longer wavelengths. Understanding whether LRDs are indeed AGNs, and, if so, how common LRDs are within the context of AGN evolution will require continued observation of larger samples of objects with JWST/NIRSpec, however. We hope that future JWST cycles will continue to find further samples of high-redshift AGNs to extend these samples. 

%\begin{acknowledgments}
\medskip
This research was funded under the JWST/NIRCam contract to the University of Arizona, NAS5-02015, and JWST Program 3215. RM, FdE, JS, and IJ acknowledge support by the Science and Technology Facilities Council (STFC), by the ERC through Advanced Grant 695671 ``QUENCH'', and by the UKRI Frontier Research grant RISEandFALL. RM also acknowledges funding from a research professorship from the Royal Society. I.J. acknowledges support by the Huo Family Foundation through a P.C. Ho PhD Studentship. H{\"U} gratefully acknowledges support by the Isaac Newton Trust and by the Kavli Foundation through a Newton-Kavli Junior Fellowship. AJB acknowledges funding from the ``FirstGalaxies'' Advanced Grant from the European Research Council (ERC) under the European Union's Horizon 2020 research and innovation programme (Grant agreement No. 789056). SC and GV acknowledges support by European Union's HE ERC Starting Grant No. 101040227 - WINGS. ECL acknowledges support of an STFC Webb Fellowship (ST/W001438/1). PGP-G acknowledges support from grant PID2022-139567NB-I00 funded by Spanish Ministerio de Ciencia e Innovaci\'on MCIN/AEI/10.13039/501100011033, FEDER, UE. MSS acknowledges support by the Science and Technology Facilities Council (STFC) grant ST/V506709/1. ST acknowledges support by the Royal Society Research Grant G125142. The research of CCW is supported by NOIRLab, which is managed by the Association of Universities for Research in Astronomy (AURA) under a cooperative agreement with the National Science Foundation.
%\end{acknowledgments}

\vspace{5mm}
\facilities{JWST(NIRCam, NIRSpec), HST(ACS)}

\software{astropy \citep{2013A&A...558A..33A,2018AJ....156..123A}}

\bibliography{ms}{}
\bibliographystyle{aasjournal}

\end{document}